\documentclass[prd,superscriptaddress,amsfonts,amssymb,amsmath,showpacs,twocolumn,nofootinbib]{revtex4-2}
\usepackage{bm}
\usepackage{amsfonts}
\usepackage{latexsym}
\usepackage{graphicx}
\usepackage{amsmath}
\usepackage{palatino}
\usepackage{xcolor} 
\usepackage{mathpazo}
\usepackage{xcolor}
\usepackage{rotating}
\usepackage{adjustbox}
\usepackage{tensor}
\usepackage{textcomp}
\usepackage{float}
\usepackage{booktabs}
\usepackage{dcolumn}
\usepackage[titletoc]{appendix}
\usepackage{booktabs}
\usepackage{multirow}
\usepackage{hyperref}
\hypersetup{colorlinks,citecolor=blue}
\usepackage{amsmath}
\usepackage{xcolor}
\usepackage{orcidlink}
\usepackage{epsfig}
\usepackage{caption}
\usepackage{subcaption}
\usepackage{commath}

\hypersetup{colorlinks,citecolor=blue}
\hypersetup{colorlinks=true,linkcolor=magenta,filecolor=magenta,    urlcolor=blue}

\def\be{\begin{equation}}
\def\ee{\end{equation}}
\def\bea{\begin{eqnarray}}
\def\eea{\end{eqnarray}}

\begin{document}

\title{Testing the Anton-Schmidt dark energy model with the DESI DR2 measurements}

\author{Himanshu Chaudhary}
\email{himanshu.chaudhary@ubbcluj.ro,\\
himanshuch1729@gmail.com}
\affiliation{Department of Physics, Babeș-Bolyai University, Kogălniceanu Street, Cluj-Napoca, 400084, Romania}
\affiliation{Research Center of Astrophysics and Cosmology, Khazar University, Baku, AZ1096, 41 Mehseti Street, Azerbaijan}

\author{Salvatore Capozziello}
\email{capozziello@na.infn.it}
\affiliation{Dipartimento di Fisica ``E. Pancini", Universit\`a di Napoli ``Federico II", Complesso Universitario di Monte Sant’ Angelo, Edificio G, Via Cinthia, I-80126, Napoli, Italy,}
\affiliation{Istituto Nazionale di Fisica Nucleare (INFN), sez. di Napoli, Via Cinthia 9, I-80126 Napoli, Italy,}
\affiliation{Scuola Superiore Meridionale, Via Mezzocannone 4, I-80134, Napoli, Italy.}

\author{Dhruba Jyoti Gogoi}
\email{moloydhruba@yahoo.in}
\affiliation{Department of Physics, Madhabdev University, Narayanpur, Lakhimpur 784164, Assam, India}

\author{G. Mustafa}
\email{gmustafa3828@gmail.com}
\affiliation{Department of Physics,
Zhejiang Normal University, Jinhua 321004, People’s Republic of China}


\begin{abstract}
We present a comparative cosmological analysis of the Anton-Schmidt, $\Lambda$CDM, and CPL models using baryon acoustic oscillation measurements from the Dark Energy Spectroscopic Instrument Data Release 2, combined with the Planck PR4 (NPIPE) CamSpec Cosmic Microwave Background likelihoods and three Type Ia supernova catalogues: Pantheon$+$, DES-Dovekie, and Union3. We use Markov Chain Monte Carlo analyses to constrain the parameters of the Anton-Schmidt model against each dataset combination. We find that the Anton-Schmidt model provides a good fit to current cosmological observations. The Anton-Schmidt model provides a good fit to the current cosmological observations but does not provide a significant alleviation of either the $H_0$ or the $S_8$ tensions. The Anton-Schmidt parameter is tightly constrained to $B \simeq -0.38$, indicating that the logarithmic correction becomes relevant only at late times. The Anton-Schmidt and CPL models predict quintessence-like present-day dark-energy equations of state and shows Quintom-B evolution, with their dark-energy equation-of-state parameters crossing the phantom divide at approximately $z \approx 0.5$. The characteristic density scale is constrained to be $\rho_\ast/\rho_{c,0}\sim4\text{--}5$, substantially smaller than the large density scales typically assumed in logotropic-inspired scenarios, indicating that the Anton-Schmidt correction mainly affects the late-time expansion history. Finally, Bayesian evidence shows moderate-to-strong preference for the Anton-Schmidt model over the $\Lambda$CDM model and strong-to-decisive preference over the CPL parametrization. Although the Anton-Schmidt model does not alleviate the current cosmological tensions, it emerges as a statistically favored dynamical dark-energy scenario whose cosmological implications deserve further tested with forthcoming Stage-IV large-scale structure observations.
\end{abstract}


\maketitle

\section{Introduction}\label{sec_1}
The $\Lambda$CDM paradigm currently stands as the cornerstone of modern cosmology, having achieved unprecedented success in describing the statistical properties of the large-scale structure (LSS), the anisotropies of the cosmic microwave background (CMB), and the background expansion history of the universe. This concordance model postulates a spatially flat universe dominated by cold dark matter (CDM) and a static cosmological constant ($\Lambda$) responsible for driving late-time cosmic acceleration \cite{Planck:2018vyg,SupernovaSearchTeam:1998fmf,Copeland:2006wr,Sahni:1999gb}. However, despite its extensive empirical validation, the $\Lambda$CDM framework is increasingly challenged by both profound theoretical puzzles and persistent observational anomalies \cite{Park:2017xbl,Perivolaropoulos:2021jda,Bullock:2017xww,Hu:2023jqc}.

Theoretically, the model is plagued by the cosmological constant problem—stemming from the extreme discrepancy between the observed value of $\Lambda$ and the vacuum energy density predicted by quantum field theory—as well as the cosmic coincidence problem \cite{Tian:2019enx,SolaPeracaula:2024nsz}. Observationally, the advent of high-precision cosmology has revealed statistically significant tensions between early-universe predictions and late-universe local measurements. Chief among these are the Hubble ($H_0$) tension and the structure growth ($S_8$) tension, which suggest that the standard model may be an incomplete description of the cosmological dynamics at low redshifts \cite{Bhattacharyya:2018fwb,Pandey:2019plg}. 

The search for new physics has been further catalyzed by the recent Dark Energy Spectroscopic Instrument (DESI) Data Release 2 (DR2). The DESI DR2 Baryon Acoustic Oscillation (BAO) measurements have shown a tantalizing statistical preference for a dynamical dark energy equation of state over a static cosmological constant. In response to these developments, the community is actively exploring alternative phenomenological frameworks, interacting dark energy scenarios, and modified gravity theories that can smoothly transition across cosmic epochs \cite{Yang:2021eud,Li:2020ybr,Giare:2024smz}.

A compelling phenomenological candidate capable of producing such dynamical behavior is the Anton-Schmidt cosmic fluid. Originally formulated to describe the equation of state of crystalline solids undergoing isotropic deformations, the Anton-Schmidt framework utilizes a logarithmic-power law relation between pressure and energy density. When adapted to a cosmological context, particularly under the Debye approximation, the Anton-Schmidt fluid offers a theoretically rich mechanism to drive late-time cosmic acceleration without explicitly invoking a fundamental cosmological constant. By modifying the effective equation of state of the dark energy fluid, this model provides a natural, thermodynamics-inspired departure from the standard $\Lambda$CDM background dynamics. 

In this work, we rigorously investigate the cosmological implications of the Anton-Schmidt cosmic fluid at the background and perturbation levels. To evaluate the observational viability of this framework, we perform a robust Markov Chain Monte Carlo (MCMC) analysis utilizing a comprehensive suite of the latest cosmological datasets. Specifically, we integrate the DESI DR2 BAO measurements with the CMB likelihoods from Planck PR4 (NPIPE), including the Planck PR4 and ACT DR6 lensing likelihoods, together with three independent Type Ia supernova compilations: the Pantheon$+$ compilation, the recalibrated DES-Dovekie sample, and the Union3 compilation.

Our primary objective is to evaluate whether the Anton-Schmidt model can act as a viable dynamical dark energy candidate and assess its capacity to alleviate the persistent $H_0$ and $S_8$ tensions. Furthermore, we aim to constrain the model's free parameters specifically the deformation parameter $B$ to determine the statistical preference of this framework relative to the standard $\Lambda$CDM paradigm using Bayesian evidence.

The manuscript is organized as follows. In Section~\ref{sec_2}, we outline the theoretical background of the Anton-Schmidt cosmological model. In Section~\ref{sec_3}, we describe the observational datasets and the MCMC methodology employed in our analysis. In Section~\ref{sec_4}, we present the observational constraints, discuss their cosmological implications, and compare the Anton-Schmidt model with the standard $\Lambda$CDM scenario using Bayesian evidence. Finally, we summarize our results and conclusions.

\section{Theoretical Background}\label{sec_2}
In standard general relativity, the interaction between gravity and matter is described by the Einstein--Hilbert action
\begin{equation}
S=\int d^4x \sqrt{-g}\left[\dfrac{R}{2}+\mathcal{L}_m(\varphi,\dot\varphi)\right],
\end{equation}
where $g$ is the determinant of the metric tensor $g_{\mu\nu}$, $R$ is the Ricci scalar, and $\mathcal{L}_m$ denotes the matter Lagrangian. Assuming the Universe to be filled with a perfect fluid, the corresponding energy-momentum tensor is given by
\begin{equation}
T_{\mu\nu}=
-\dfrac{2}{\sqrt{-g}}
\dfrac{\delta \mathcal{L}_m}{\delta g^{\mu\nu}}.
\end{equation}

Following the scalar-field formulation discussed in \cite{Barrow:1990vx}, the energy density and pressure can be written as
\begin{align}
\epsilon_\varphi&=
\dfrac{1}{2}\dot{\varphi}^2+V(\varphi),
\label{eq:rho_fi}
\\
P_\varphi&=
\dfrac{1}{2}\dot{\varphi}^2-V(\varphi),
\label{eq:p_fi}
\end{align}
where $\varphi$ is the scalar field and $V(\varphi)$ its self-interacting potential.

We consider a cosmological fluid described by the Anton--Schmidt equation of state, originally motivated by the behavior of crystalline solids under isotropic deformations \cite{Capozziello:2017buj,Capozziello:2018mds},
\begin{equation}
P=
A
\left(
\dfrac{\epsilon}{\epsilon_\ast}
\right)^{-n}
\ln
\left(
\dfrac{\epsilon}{\epsilon_\ast}
\right),
\label{eq:Anton-Schmidt}
\end{equation}
where $\epsilon_\ast$ is a reference energy density and $n$ is related to the Gr\"uneisen parameter $\gamma_G$ through $n=-1/6-\gamma_G$ \cite{Gruneisen1912}. In contrast to the original formulation, where the equation of state is expressed in terms of the rest-mass density $\rho$, here $\epsilon$ denotes the total energy density of the cosmic fluid. The Anton--Schmidt cosmological model can therefore be interpreted as an effective scalar-field scenario described by the Lagrangian
\begin{equation}
\mathcal{L}_m=
\mathcal{K}(\dot{\varphi})-V(\varphi),
\end{equation}
where the kinetic term is assumed to take the canonical form $\mathcal{K}(\dot{\varphi})=\dot{\varphi}^2/2$.

Assuming a homogeneous and isotropic spatially flat FLRW universe,
\begin{equation}
ds^2=
dt^2-a(t)^2
\left[
dr^2+r^2(d\theta^2+\sin^2\theta\,d\phi^2)
\right],
\end{equation}
the Friedmann and continuity equations become
\begin{align}
H^2&=
\left(
\dfrac{\dot a}{a}
\right)^2
=
\dfrac{\epsilon}{3},
\label{eq:Friedmann}
\\
\dot{\epsilon}
+
3H(\epsilon+P)
&=0,
\label{eq:continuity}
\end{align}
where $a(t)$ is the scale factor normalized to unity at the present epoch, while $\epsilon$ and $P$ denote the total energy density and pressure of the Anton--Schmidt fluid, respectively.
\subsection{The Anton--Schmidt Cosmological Model}
\label{sec_2b}
The Anton-Schmidt fluid in the Debye approximation provides a phenomenological mechanism capable of explaining the late-time accelerated expansion of the Universe without explicitly introducing a cosmological constant \cite{Capozziello:2017buj}. As the Universe evolves, the Anton--Schmidt equation of state naturally allows transitions between decelerating and accelerating phases. In the present work, we focus on the special case $n=-1$, which provides a viable description of the late-time cosmic dynamics at the background level. The motivation for this choice and its physical implications will be analyzed in detail later in this section.

To investigate the cosmological implications of this scenario, we consider the original Anton--Schmidt equation of state
\begin{equation}
P=A\left(\dfrac{\rho}{\rho_\ast}\right)
\ln \left(\dfrac{\rho}{\rho_\ast}\right),
\label{eq:AS_pressure}
\end{equation}
where $A$ is a constant, $\rho_\ast$ is a reference density, and $\rho$ denotes the rest-mass density. We assume a spatially flat FLRW universe filled with a perfect fluid of total energy density $\epsilon$.

Under the assumption of adiabatic evolution, the first law of thermodynamics reads
\begin{equation}
d\epsilon=
\left(
\dfrac{\epsilon+P}{\rho}
\right)d\rho.
\label{eq:AS_firstlaw}
\end{equation}
Integrating Eq.~\eqref{eq:AS_firstlaw}, one obtains
\begin{equation}
\epsilon=
\rho+
\rho
\int^\rho
d\rho'
\dfrac{P(\rho')}{\rho'^2}.
\label{eq:AS_integral}
\end{equation}

Substituting Eq.~\eqref{eq:AS_pressure} into Eq.~\eqref{eq:AS_integral}, the total energy density becomes
\begin{equation}
\epsilon=
\rho+
\dfrac{A}{2}
\left(
\dfrac{\rho}{\rho_\ast}
\right)
\ln^2
\left(
\dfrac{\rho}{\rho_\ast}
\right).
\label{eq:AS_total_density}
\end{equation}

The case $n=-1$ corresponds to a distinct analytical limit of the Anton--Schmidt equation of state and cannot be directly obtained from the general $n\neq-1$ formulation discussed in \cite{Capozziello:2017buj}. Consequently, the $n=-1$ scenario should be regarded as an independent cosmological realization, whose physical implications may differ from those of pure logotropic models \cite{Chavanis:2016pcp}. In this framework, the first term in Eq.~\eqref{eq:AS_total_density} dominates at early times $(a\ll1)$, reproducing an effective matter-dominated epoch, while at late times $(a\gg1)$ the pressure becomes negative, leading to a dark-energy-dominated accelerated expansion. This naturally motivates decomposing the total energy density into matter and dark-energy contributions, namely $\epsilon=\epsilon_m+\epsilon_{de}$, where the matter and dark-energy components are respectively given by

\begin{align}
&\epsilon_m=\dfrac{\rho_{m,0}}{a^3}\ , \label{eq:rho_m}\\
&\epsilon_{de}=\dfrac{\epsilon_{de,0} }{a^3}-\dfrac{3A}{a^3}\left(\dfrac{\rho_{m,0}}{\rho_*}\right)\ln a \ln \left(\dfrac{\rho_{m,0}}{\rho_*}a^{-3/2}\right),
\label{eq:rho_de}
\end{align}
where
\begin{equation}
\epsilon_{de,0}=\dfrac{A}{2}\left(\dfrac{\rho_{m,0}}{\rho_\ast}\right)\ln^2\left(\dfrac{\rho_{m,0}}{\rho_\ast}\right).
\label{eq:rho_de0}
\end{equation}

Defining the normalized density parameters
\begin{equation}
\Omega_{m,0}\equiv
\dfrac{\epsilon_{m,0}}{\epsilon_{c,0}},
\qquad
\Omega_{de,0}\equiv
\dfrac{\epsilon_{de,0}}{\epsilon_{c,0}}
=
1-\Omega_{m,0},
\label{eq:AS_omegas}
\end{equation}
where $\epsilon_{c,0}=3H_0^2$ is the present critical density, the Friedmann equation becomes
\begin{equation}
\begin{split}
H^2(a)=H_0^2\Bigg[
&\frac{\Omega_{m,0}}{a^3}
+\frac{1-\Omega_{m,0}}{a^3} \\
&\times \left(1-6B\ln a+9B^2\ln^2 a\right)
\Bigg].
\end{split}
\end{equation}

where we define the parameter
\begin{equation}
B\equiv
\ln^{-1}
\left(
\dfrac{\rho_{m,0}}{\rho_\ast}
\right).
\label{eq:B_def}
\end{equation}

It is worth noting that the parameter $B$ introduced in Eq.~\eqref{eq:B_def} differs from the corresponding quantity appearing in pure logotropic models. In the logotropic scenario, the analogous parameter is interpreted as the dimensionless logotropic temperature and is defined as \cite{Chavanis:2016pcp}
\begin{equation}
B_{\rm log}\equiv
\left[
\ln\left(
\dfrac{\rho_\ast}{\rho_{m,0}}
\right)-1
\right]^{-1}.
\end{equation}
Assuming $\rho_\ast$ to be of the order of the Planck density, one generally obtains $0<B_{\rm log}\ll1$ \cite{Chavanis:2015paa,Chavanis:2015eka,Chavanis:2016pcp}. In contrast, within the present Anton--Schmidt framework, the condition $\rho_\ast\gg1$ naturally leads to negative values of $B$.

Using Eq.~\eqref{eq:B_def}, the total energy density can be rewritten as
\begin{equation}
\dfrac{\epsilon}{\epsilon_{c,0}}
=
\dfrac{\Omega_{m,0}}{a^3}
+
\dfrac{1-\Omega_{m,0}}{a^3}
\left(
1-6B\ln a+9B^2\ln^2 a
\right).
\label{eq:AS_density_final}
\end{equation}

and
\begin{align}
P&=
\dfrac{A}{a^3}
\left(
\dfrac{\rho_{m,0}}{\rho_\ast}
\right)
\left[
\ln\left(
\dfrac{\rho_{m,0}}{\rho_\ast}
\right)
-3\ln a
\right]
\nonumber\\
&=
\dfrac{A}{a^3}
e^{1/B}
\left(
\dfrac{1}{B}-3\ln a
\right).
\label{eq:AS_pressure_intermediate}
\end{align}

Using Eq.~\eqref{eq:rho_de0}, one obtains
\begin{equation}
\epsilon_{de,0}=
\dfrac{A}{2}
\dfrac{e^{1/B}}{B^2},
\label{eq:AS_de0_B}
\end{equation}
which allows the pressure to be rewritten as
\begin{equation}
P=
2\epsilon_{c,0}
\dfrac{1-\Omega_{m,0}}{a^3}
\left(
B-3B^2\ln a
\right).
\label{eq:P}
\end{equation}

The total equation-of-state parameter is therefore given by
\begin{equation}
w(a)=
\dfrac{
2B-6B^2\ln a
}{
(1-\Omega_{m,0})^{-1}
-6B\ln a
+9B^2\ln^2 a
}.
\label{eq:AS_total_w}
\end{equation}

Since $B<0$ and $\Omega_{m,0}<1$, the effective equation-of-state parameter remains negative throughout the cosmic evolution and asymptotically approaches zero from below at very large scale factors. In particular, at the present epoch $(a=1)$ one finds
\begin{equation}
w_0=
2B(1-\Omega_{m,0}).
\label{eq:AS_w0_total}
\end{equation}

Identifying the total pressure with the effective dark-energy pressure \cite{Capozziello:2017buj}, the dark-energy equation-of-state parameter becomes
\begin{equation}
w_{de}=
\dfrac{P}{\epsilon_{de}}
=
\dfrac{2B}{1-3B\ln a}.
\label{eq:AS_wde}
\end{equation}

We note that the effective dark-energy equation of state formally diverges when
\begin{equation}
1-3B\ln a=0.
\end{equation}
However, this behavior does not correspond to a physical singularity of the cosmological background evolution, since the Hubble expansion rate remains finite. The divergence arises because the effective dark-energy density vanishes while the pressure remains finite, indicating a breakdown of the effective fluid decomposition rather than a true spacetime singularity.

At the present epoch, Eq.~\eqref{eq:AS_wde} reduces to
\begin{equation}
w_{de,0}=2B.
\label{eq:AS_wde0}
\end{equation}

The $\Lambda$CDM limit is recovered for $B\rightarrow -1/2$ at $a=1$. Expanding Eq.~\eqref{eq:AS_wde} around the present epoch yields
\begin{equation}
w_{de}\approx
2B+6B^2(a-1),
\label{eq:AS_wde_approx}
\end{equation}
showing that the Anton-Schmidt model naturally behaves as a dynamical dark-energy scenario. This behavior is analogous to the Chevallier-Polarski-Linder (CPL) parametrization \cite{Chevallier:2000qy,Linder:2002et,Linder:2024rdj}, although in the present case the parameters are not independent but are both determined by the single parameter $B$.

An important quantity characterizing the perturbative behavior of an adiabatic fluid is the sound speed \cite{Mukhanov:1990me}, defined as
\begin{equation}
c_a^2 \equiv \frac{\partial P}{\partial \epsilon}.
\label{eq:ca}
\end{equation}

Using Eqs.~\eqref{eq:AS_pressure} and \eqref{eq:AS_total_density}, we obtains
\begin{align}
c_a^2&=
\left(
\frac{\partial P}{\partial\rho}
\right)
\left(
\frac{\partial\epsilon}{\partial\rho}
\right)^{-1}\\
&=
\frac{
A\left[1+\ln\left(\rho/\rho_\ast\right)\right]
}{
\rho_\ast+
\dfrac{A}{2}
\left[
2+\ln\left(\rho/\rho_\ast\right)
\right]
\ln\left(\rho/\rho_\ast\right)
}.
\label{eq:ca_rho}
\end{align}

In the matter-dominated epoch, where $\epsilon\simeq\rho$, we can calculate the adiabatic sound speed for the Anton-Schmidt fluid in terms of the parameter $B$ by using
\begin{align}
\frac{\partial \rho}{\partial a}
&=
-\frac{3\epsilon_{c,0}\Omega_{m,0}}{a^4},
\\
\frac{\partial P}{\partial a}
&=
-\frac{6\epsilon_{c,0}(1-\Omega_{m,0})}{a^4}
B\left(1+B-3B\ln a\right).
\end{align}
Consequently, the adiabatic sound speed can be expressed as
\begin{equation}
c_a^2=
\left(\frac{\partial \rho}{\partial a}\right)^{-1}
\frac{\partial P}{\partial a}
=
\frac{2B(1-\Omega_{m,0})\left(1+B-3B\ln a\right)}
{\Omega_{m,0}}.
\label{eq:cs_nuovo}
\end{equation}

Since the parameter $B$ is defined differently in the Anton-Schmidt and pure logotropic models, the corresponding sound speeds are related through
\begin{equation}
c_a=
\sqrt{
\frac{2(1+B)\left(3B\ln a-B-1\right)}
{a^3}
}
\,c_{s,\mathrm{log}},
\label{eq:comparison}
\end{equation}
where $B$ is defined in Eq.~\eqref{eq:B_def} and $c_{s,\mathrm{log}}$ denotes the sound speed of the pure logotropic model. We emphasize that, at the level of linear perturbations, the parameter $n$ becomes a function of the temperature. Consequently, the assumption $n=-1$ cannot be expected to remain valid throughout the entire expansion history of the Universe, and the corresponding expression for the adiabatic sound speed given by Eq.~\eqref{eq:cs_nuovo} may not be applicable over the full redshift range. Nevertheless, this does not prevent the process of structure formation nor significantly affect its overall dynamics, indicating that the Anton-Schmidt model remains a viable description of the early-time cosmological evolution \cite{Capozziello:2017buj}.

In the present work, we focus on the special case $n=-1$, which should be regarded as an effective description of the late-time Universe rather than a fundamental value valid throughout the entire cosmic history. The Anton--Schmidt equation of state is derived within the Debye approximation, whose validity is expected to break down at sufficiently high temperatures. Consequently, the parameter $n$, related to the Gr\"uneisen parameter through
\begin{equation}
n=-\frac{1}{6}-\gamma_G,
\end{equation}
is generally expected to depend on temperature rather than remain constant throughout the cosmic evolution.

Following~\cite{nie2010temperature}, the temperature dependence of the Gr\"uneisen parameter can be written as
\begin{equation}
\gamma_G(T)=\gamma_{G,0}\left[1+b_1(T-T_0)+b_2(T-T_0)^2\right],
\label{eq:gamma_T}
\end{equation}
where $T_0$ denotes the reference temperature, while $b_1=1.45\times10^{-4}\,\mathrm{K}^{-1}$ and $b_2=5.40\times10^{-7}\,\mathrm{K}^{-2}$ are experimentally determined coefficients~\cite{nie2010temperature}. Consequently, the Anton-Schmidt parameter becomes
\begin{equation}
n(T)=-\frac{1}{6}-\gamma_G(T).
\label{eq:n_T}
\end{equation}

\begin{figure}
\centering
\includegraphics[scale=0.30]{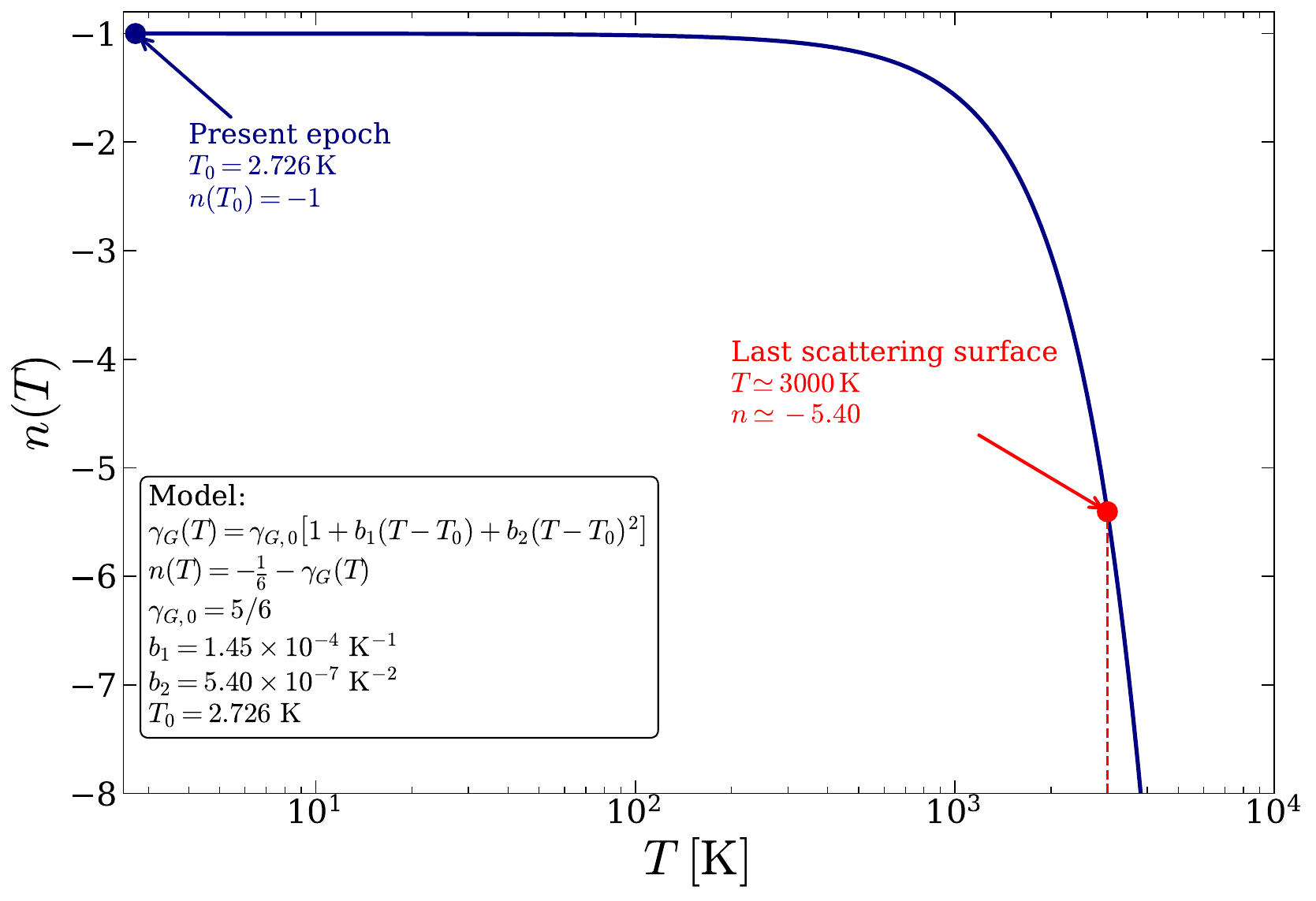}
\caption{Evolution of the Anton--Schmidt parameter $n$ as a function of temperature, obtained from Eqs.~\eqref{eq:gamma_T} and \eqref{eq:n_T}. The present work adopts the late-time limit $n=-1$, corresponding to the current CMB temperature $T_0=2.726\,\mathrm{K}$.}\label{fig_1}
\end{figure}

Throughout this work, we fix $\gamma_{G,0}=5/6$, which gives $n(T_0)=-1$ at the present epoch. The resulting evolution of $n(T)$ is shown in Fig.~\ref{fig_1}. As the temperature increases, the parameter gradually changes from its present-day value. In particular, at the last-scattering epoch ($T\simeq3000\,\mathrm{K}$), we find
\begin{equation}
n_{\rm CMB}\simeq -5.40.
\end{equation}

The above arguments shows that the choice $n=-1$ is not a theoretically preferred value of the general Anton-Schmidt equation of state, but rather an effective low-temperature (late-time) limit of a more general temperature-dependent Anton--Schmidt fluid. Therefore, nearby values of $n$ are not excluded by the present analysis, nor are they ruled out by current observations. Instead, they belong to the broader Anton--Schmidt family and would require an independent cosmological analysis. Accordingly, our analysis is restricted to the constant $n=-1$ realization, and the CAMB Boltzmann solver is modified consistently under this assumption. Extending the analysis to the full temperature-dependent Anton--Schmidt model would require implementing the evolution of $n(T)$ in both the background and perturbation equations of the Boltzmann solver, followed by a new cosmological parameter estimation. Such an investigation is beyond the scope of this work and is left for future study.

\begin{figure}
\centering
\includegraphics[scale=0.5]{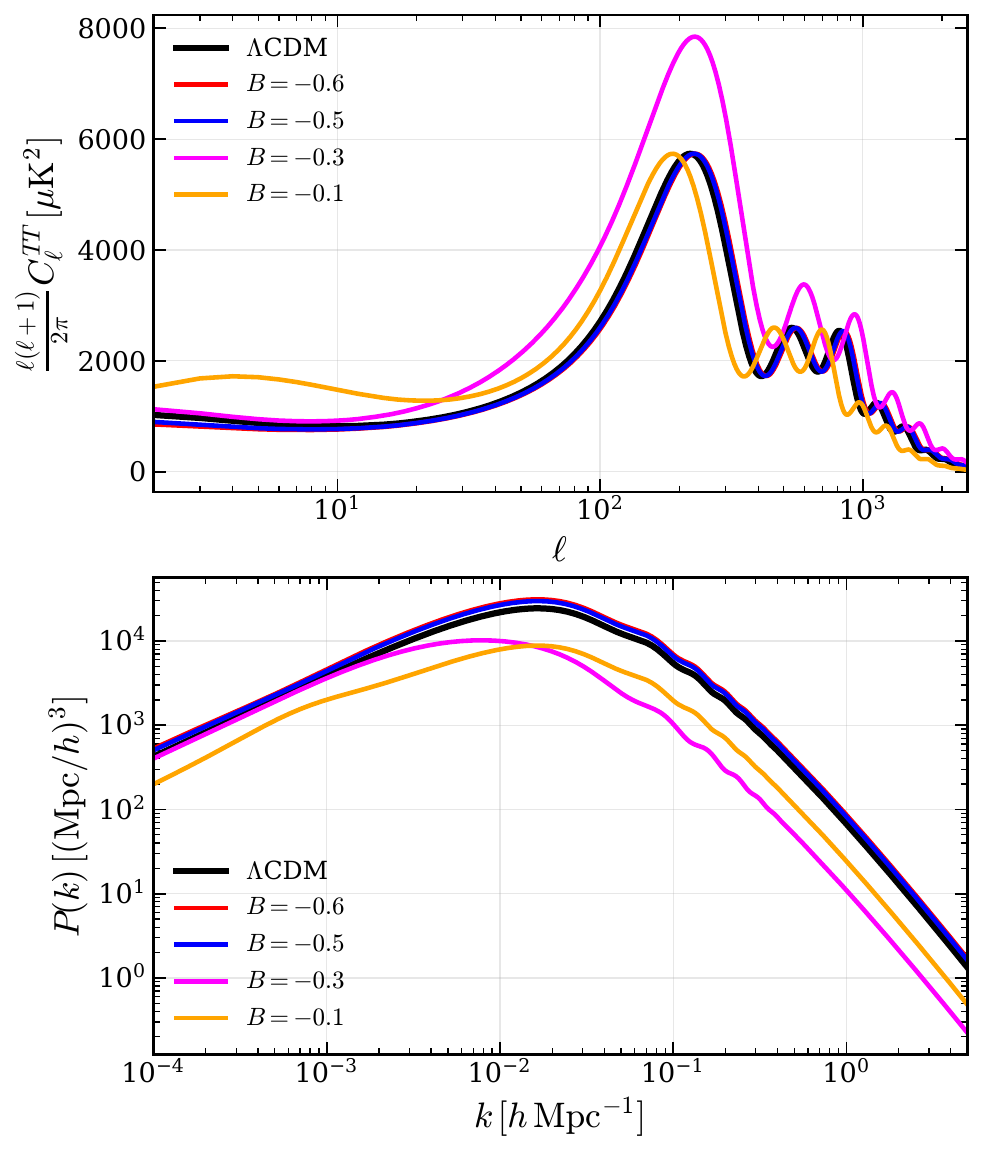}
\caption{The CMB TT angular power spectrum (upper panel) and the matter power spectrum (lower panel) for different values of the Anton--Schmidt parameter ($B$)}\label{fig_new}
\end{figure}

Fig~\ref{fig_new} shows the CMB TT angular power spectrum (upper panel) and the linear matter power spectrum (lower panel) for different values of the Anton-Schmidt parameter $B$. The upper panel shows that the positions and amplitudes of the acoustic peaks depend sensitively on the value of $B$. Models with $B=-0.6$ and $B=-0.5$ remain close to the $\Lambda$CDM prediction over most multipoles, indicating only mild modifications to the expansion history prior to recombination. In contrast, larger values such as $B=-0.3$ and $B=-0.1$ produce substantial changes in both the height and shape of the acoustic peaks, leading to increasingly large deviations from the $\Lambda$CDM spectrum.

The lower panel shows that the growth of matter perturbations is also significantly affected by the value of $B$. While the models with $B=-0.6$ and $B=-0.5$ closely follow the $\Lambda$CDM matter power spectrum over a wide range of scales, larger values of $B$ suppress the amplitude of the matter power spectrum, particularly on intermediate and small scales ($k \gtrsim 10^{-2}\,h\,\mathrm{Mpc}^{-1}$). The suppression becomes progressively stronger as $B$ approaches zero.

\subsection{The Chevallier-Polarski-Linder Parametrization}
To facilitate a direct comparison between the Anton-Schmidt framework and standard dynamical dark energy scenarios, we compare our results against the well-known Chevallier-Polarski-Linder (CPL) parametrization \cite{Chevallier:2000qy,Linder:2002et,Linder:2024rdj}. In this phenomenological model, the effective dark-energy equation-of-state parameter is expanded to first order in the scale factor $a$, taking the form
\begin{equation}
w_{de}(a)=w_0+w_a(1-a),
\end{equation}
where $w_0$ represents the present-day value of the equation of state, and $w_a$ characterizes its dynamical evolution over cosmic time. 
 
By substituting this equation of state into the standard continuity equation, the evolution of the dark-energy density for the CPL parametrization is obtained as
\begin{equation}
\epsilon_{de}(a)=\epsilon_{de,0}a^{-3(1+w_0+w_a)}\exp\left[-3w_a(1-a)\right].
\end{equation}
 
Assuming a spatially flat FLRW background composed of matter and CPL dark energy, the corresponding Friedmann equation governing the background expansion history becomes
\begin{equation}
\begin{split}
H^2(a) &= H_0^2 \bigg[
\frac{\Omega_{m,0}}{a^3} \\
&+ (1-\Omega_{m,0})a^{-3(1+w_0+w_a)}
\exp\left[-3w_a(1-a)\right]
\bigg].
\end{split}
\end{equation}
While the CPL model introduces $w_0$ and $w_a$ as two independent free parameters to describe the dynamical nature of dark energy, the Anton-Schmidt model captures a similarly rich dynamical behavior (as shown in the phantom-crossing evolution in Eq.~\eqref{eq:AS_wde_approx}) using only a single deformation parameter $B$.

\section{Dataset and Methodology}\label{sec_3}
To constrain the parameters of the Anton-Schmidt cosmological model, we perform a Markov Chain Monte Carlo (MCMC) analysis using the cosmological inference package {\tt Cobaya}\footnote{\url{https://github.com/CobayaSampler/cobaya}} \cite{torrado2021cobaya}. The exploration of the parameter space is carried out through the Metropolis-Hastings sampling algorithm~\cite{lewis2002cosmological,lewis2013efficient,neal2005taking}, while the convergence of the chains is monitored using the Gelman-Rubin criterion, requiring $R-1<0.01$~\cite{gelman1992inference}. The theoretical predictions are computed with the Boltzmann solver {\tt CAMB}\footnote{\url{https://github.com/cmbant/CAMB.git}} \cite{lewis2000efficient,howlett2012cmb}, and the resulting chains are analyzed using the {\tt GetDist}\footnote{\url{https://github.com/cmbant/getdist}} package \cite{lewis2025getdist} for statistical post-processing and visualization.

In our analysis, we use the combination of late-and early-universe observations, including Baryon Acoustic Oscillation measurements from the Dark Energy Spectroscopic Instrument Data Release 2, Type Ia supernova compilations from Pantheon${+}$, DES-Dovekie, and Union3, together with the CamSpec Cosmic Microwave Background likelihood based on the latest Planck NPIPE PR4 release. The details of these datasets are summarized below.
\begin{itemize}
     \item \textbf{Baryon Acoustic Oscillation :} We use the BAO measurements from the DESI DR2\footnote{\url{https://github.com/CobayaSampler/bao_data.git}} \cite{karim2025desi}, obtained from multiple tracers including BGS, LRG, ELG, QSO, and Lyman-$\alpha$ forests. In our analysis, we compute the Hubble distance $D_H(z)=c/H(z)$, the comoving angular diameter distance $D_M(z)=c\int_0^z \frac{dz'}{H(z')}$, and the volume-averaged distance $D_V(z)=\left[zD_M^2(z)D_H(z)\right]^{1/3}$. Using these quantities, we evaluate the BAO observables $D_M/r_d$, $D_H/r_d$, $D_V/r_d$, and $D_M/D_H$, where $r_d$ denotes the sound horizon at the drag epoch.
     
     \item \textbf{Type Ia Supernova :} We further consider three different Type Ia supernova (SNe Ia) compilations. First, we use the Pantheon${+}$ sample~\footnote{\url{https://github.com/PantheonPlusSH0ES/DataRelease.git}} \cite{scolnic2022pantheon}, consisting of 1,701 light curves from 1,550 SNe Ia spanning the redshift range $0.001 \leq z \leq 2.26$. In our analysis, we exclude data with $z<0.01$ to reduce systematic uncertainties associated with peculiar velocities \cite{brout2022pantheon}. Second, we use the DES-Dovekie sample~\footnote{\url{https://github.com/des-science/DES-SN5YR.git}} \cite{popovic2025dark}, which contains 1,820 recalibrated photometric light curves in the range $0.025<z<1.14$, including 1,623 DES-discovered SNe Ia and 197 low-$z$ supernovae from the CfA and CSP compilations \cite{hicken2009cfa3,hicken2012cfa4,foley2017foundation}. Finally, we include the Union3~\footnote{\url{https://github.com/rubind/union3_release.git}} datasets \cite{rubin2025union}, comprising 2,087 SNe Ia over the interval $0.01<z<2.26$, with 1,363 supernovae overlapping with the Pantheon${+}$ sample.
     
     \item \textbf{CamSpec CMB likelihood :} Finally, we use a combination of CMB datasets. Our analysis includes the Planck temperature \texttt{(TT)}, polarization \texttt{(EE)}, and temperature–polarization cross-correlation \texttt{(TE)} power spectra, using the low-$\ell$ \texttt{Commander} and \texttt{simall} likelihoods ($\ell < 30$) together with the high-$\ell$ \texttt{CamSpec} likelihood ($\ell \geq 30$) from the latest Planck PR4 (NPIPE) release~\footnote{\url{https://github.com/carronj/planck_PR4_lensing.git}}\cite{Efstathiou2021Detailed,rosenberg2022cmb}. We further include the Planck PR4 lensing likelihood \cite{carron2022cmb,carron2022planck} and the ACT DR6 lensing likelihood~\footnote{\url{https://github.com/ACTCollaboration/act_dr6_lenslike.git}} \cite{madhavacheril2024atacama,qu2024atacama}.
\end{itemize}

To compare the Anton--Schmidt model with the standard $\Lambda$CDM cosmology, we perform a Bayesian model comparison based on the Bayesian evidence (or marginal likelihood). For a cosmological model $M$ with parameter vector $\boldsymbol{\Theta}$ and observational data $D$, the Bayesian evidence is defined as
\begin{equation}
\mathcal{Z}(M)\equiv P(D|M)
=
\int
\mathcal{L}(D|\boldsymbol{\Theta},M)\,
\pi(\boldsymbol{\Theta}|M)\,
d\boldsymbol{\Theta},
\end{equation}
where $\mathcal{L}(D|\boldsymbol{\Theta},M)$ is the likelihood function and $\pi(\boldsymbol{\Theta}|M)$ denotes the prior distribution of the model parameters. The Bayesian evidence is obtained by integrating the likelihood over the full prior volume and therefore naturally accounts for both the goodness of fit and the complexity of the model.

The relative preference between two competing cosmological models, $M_i$ and $M_j$, is determined through Bayes' theorem,
\begin{equation}
\frac{P(M_i|D)}{P(M_j|D)}
=
\frac{P(D|M_i)}{P(D|M_j)}
\,
\frac{P(M_i)}{P(M_j)},
\end{equation}
where $P(M_i)$ and $P(M_j)$ are the prior probabilities assigned to the models. The ratio of Bayesian evidences,
\begin{equation}
\mathcal{B}_{ij}
=
\frac{\mathcal{Z}_i}{\mathcal{Z}_j},
\end{equation}
defines the Bayes factor, which quantifies the relative support of the data for one model over another. In this work we quote the logarithmic Bayes factor,
\begin{equation}
\ln\mathcal{B}_{ij}
=
\ln\mathcal{Z}_i-\ln\mathcal{Z}_j.
\end{equation}
A positive value of $\ln\mathcal{B}_{ij}$ indicates a preference for model $i$, whereas a negative value favours model $j$. To interpret the strength of the evidence, we adopt the revised Jeffreys' scale \cite{kass1995bayes,trotta2008bayes}: $\ln\mathcal{B}_{ij}<1$ corresponds to inconclusive evidence, $1\leq\ln\mathcal{B}_{ij}<2.5$ to weak evidence, $2.5\leq\ln\mathcal{B}_{ij}<5$ to moderate evidence, $5\leq\ln\mathcal{B}_{ij}<10$ to strong evidence, and $\ln\mathcal{B}_{ij}\geq10$ to decisive evidence.

The Bayesian evidence is computed using the \texttt{MCEvidence} package~\footnote{\url{https://github.com/yabebalFantaye/MCEvidence.git}} \cite{heavens2017marginal,heavens2017no}, implemented through the Cobaya wrapper available in the \texttt{wgcosmo} repository~\footnote{\url{https://github.com/williamgiare/wgcosmo.git}}. In the present analysis, we use physically motivated priors for all cosmological parameters, as summarized in Table~\ref{tab_1}. The \texttt{MCEvidence} algorithm estimates the Bayesian evidence directly from converged MCMC chains using a nearest-neighbour density estimator. Since the Bayesian evidence is obtained by integrating the likelihood over the adopted prior volume, the resulting Bayes factors may depend on the choice of priors. Bayesian evidence is determined by the integrated likelihood over the full multidimensional parameter space rather than by visual differences between the best-fit predictions alone. Consequently, even relatively small differences in the likelihood and the corresponding posterior volume may lead to noticeable differences in the Bayesian evidence.
\begin{table}[t]
\centering
\renewcommand{\arraystretch}{1}
\begin{tabular}{|lll|}
\hline
Models & parameter & prior\\  
\hline 
$\mathbf{\Lambda}$\textbf{CDM} & $\Omega_\mathrm{cdm}h^2$ & $\mathcal{U}[0.001, 0.99]$ \\   
& $\Omega_\mathrm{b}h^{2}$ & $\mathcal{U}[0.005, 0.1]$ \\
& $100\theta_\mathrm{MC}$ & $\mathcal{U}[0.5, 10]$ \\
& $n_\mathrm{s}$ & $\mathcal{U}[0.8, 1.2]$ \\
& $\tau_{\mathrm{reio}}$ & $\mathcal{U}[0.01, 0.8]$ \\
& $\ln(10^{10} A_\mathrm{s})$ & $\mathcal{U}[1.61, 3.91]$ \\
\hline 
\textbf{Dark energy} & $w_0$ & $\mathcal{U}[-3, 1]$ \\
& $w_{a}$ & $\mathcal{U}[-3, 2]$ \\
\hline 
\textbf{Anton Schmidt} & $B$ & $\mathcal{U}[-1, 0]$ \\
\hline
\end{tabular}
\caption{Parameters and priors used in the analysis. Here $\mathcal{U}[{\rm min, max}]$ denotes a uniform prior over the specified range.}\label{tab_1}
\end{table}

\begin{figure*}
\centering
\includegraphics[scale=0.42]{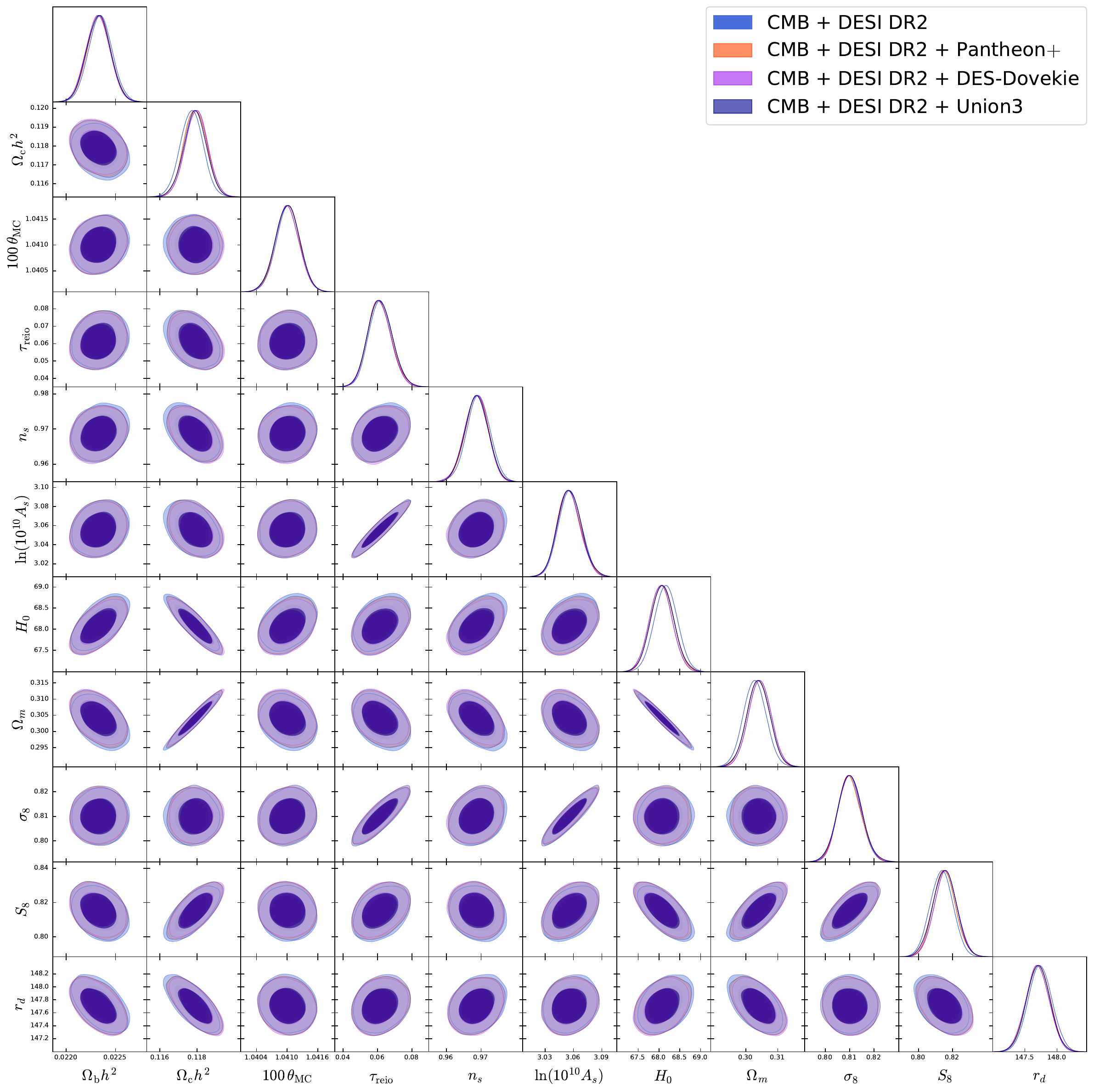}
\caption{The figure shows the corner plot of the $\Lambda$CDM model at 68\% ($1\sigma$) and 95\% ($2\sigma$) confidence levels using DESI DR2 combined with CMB and SNe~Ia datasets (Pantheon$+$, DES-Dovekie, and Union3), shown as superimposed contours for the different dataset combinations.}\label{fig_2}
\end{figure*}
\begin{figure*}
\centering
\includegraphics[scale=0.42]{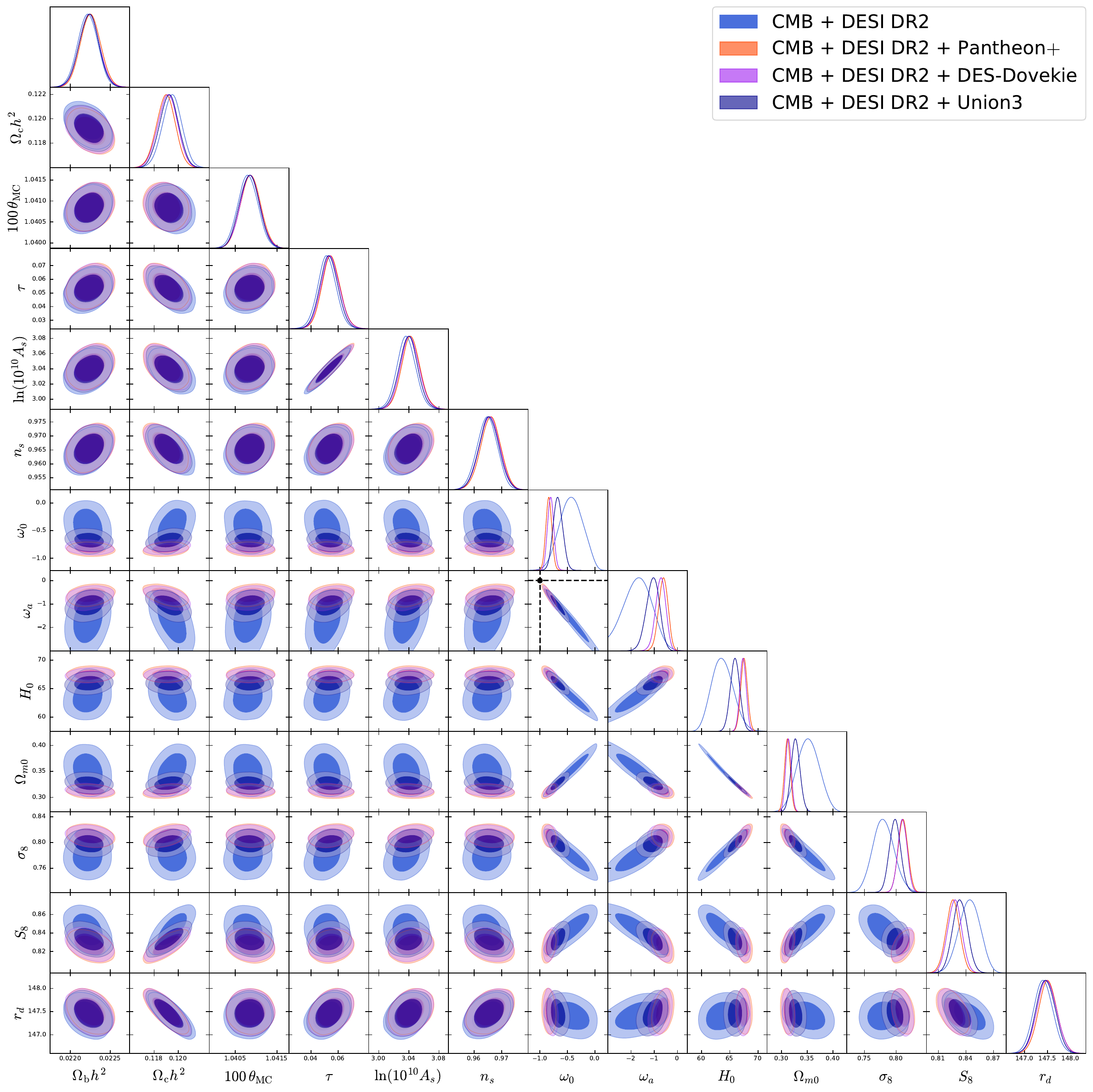}
\caption{The figure shows the corner plot of the CPL model at 68\% ($1\sigma$) and 95\% ($2\sigma$) confidence levels using DESI DR2 combined with CMB and SNe~Ia datasets (Pantheon$+$, DES-Dovekie, and Union3), shown as superimposed contours for the different dataset combinations.}\label{fig_3}
\end{figure*}
\begin{figure*}
\centering
\includegraphics[scale=0.42]{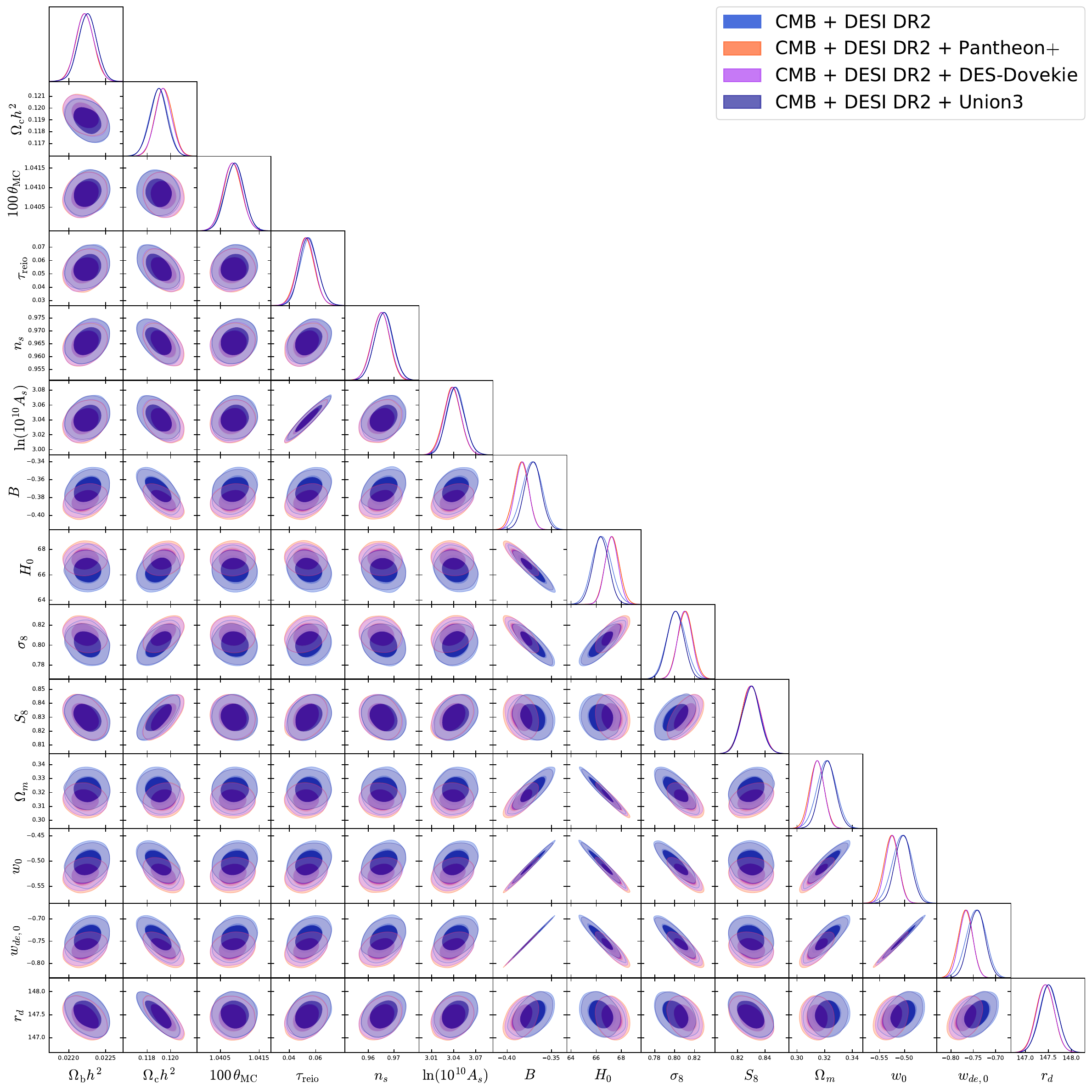}
\caption{The figure shows the corner plot of the Anton-Schmidt model at 68\% ($1\sigma$) and 95\% ($2\sigma$) confidence levels using DESI DR2 combined with CMB and SNe~Ia datasets (Pantheon$+$, DES-Dovekie, and Union3), shown as superimposed contours for the different dataset combinations.}\label{fig_4}
\end{figure*}

\begin{figure*}
\begin{subfigure}{0.45\textwidth}
\includegraphics[width=\linewidth]
{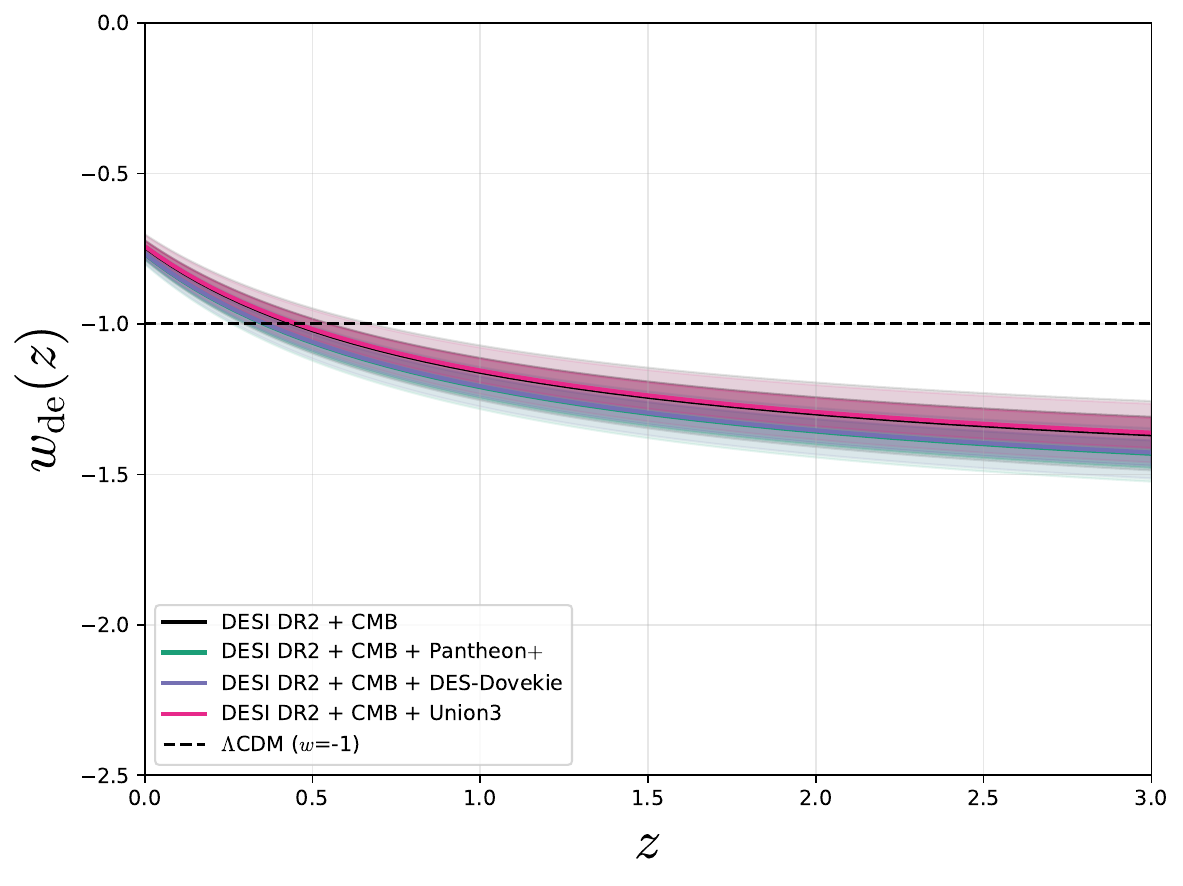}
\end{subfigure}
\hfil
\begin{subfigure}{0.45\textwidth}
\includegraphics[width=\linewidth]
{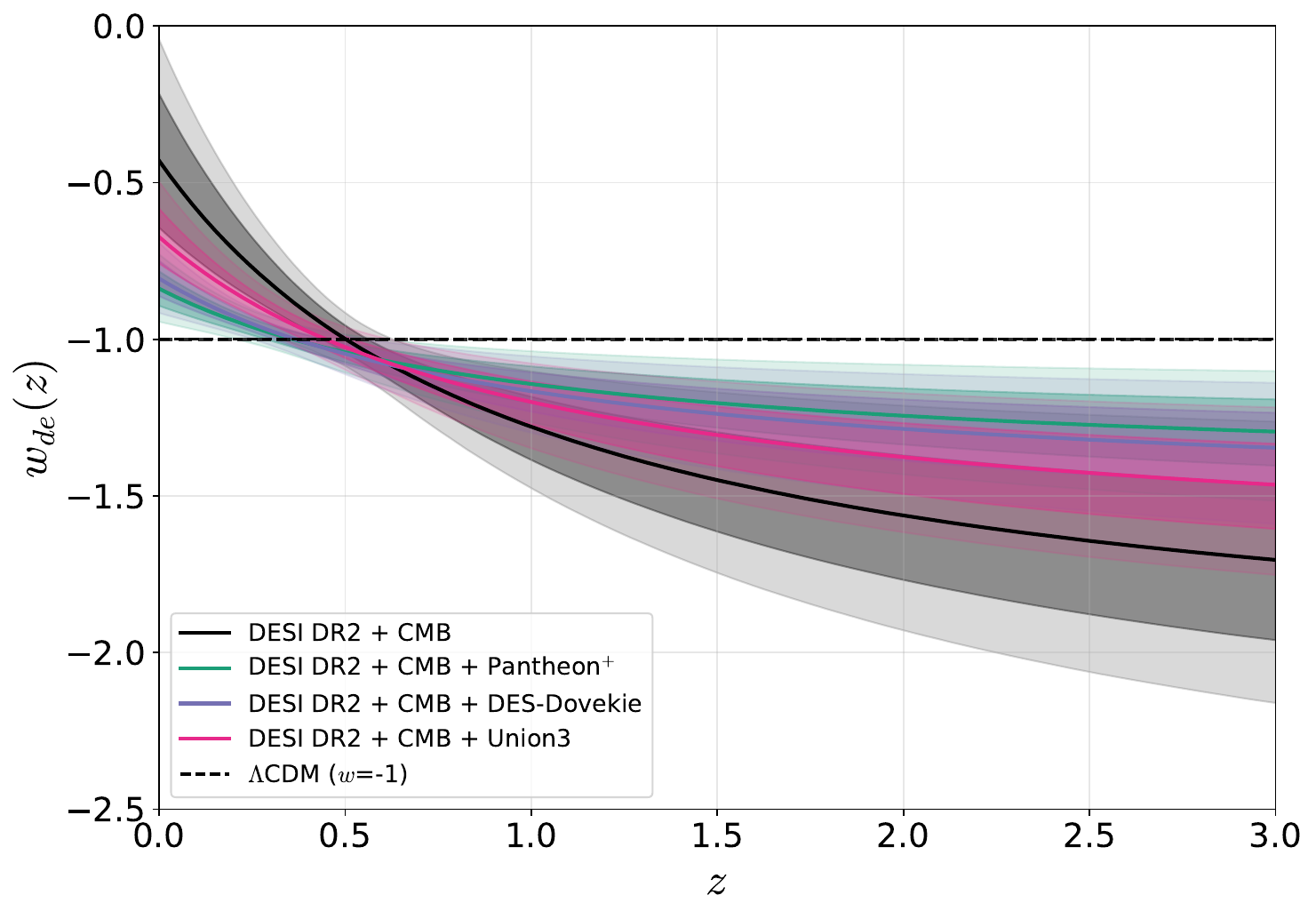}
\end{subfigure}
\caption{This figure shows the evolution of the effective dark-energy equation-of-state parameter, $w_{de}(z)$, for the Anton-Schmidt model (left panel) and the CPL model (right panel) using DESI DR2 combined with CMB and different SNe~Ia datasets. The solid lines denote the mean evolution, while the dark and light shaded regions correspond to the 68\% ($1\sigma$) and 95\% ($2\sigma$) confidence intervals, respectively.}\label{fig_5}
\end{figure*}

\begin{table*}
\centering
\resizebox{\textwidth}{!}{%
\begin{tabular}{l@{\hspace{10pt}}c@{\hspace{10pt}}c@{\hspace{10pt}}c@{\hspace{10pt}}c}
\hline
\textbf{Parameters} 
& \textbf{CMB + DESI DR2} 
& \textbf{CMB + DESI DR2 + Pantheon$+$} 
& \textbf{CMB + DESI DR2 + DES-Dovekie} 
& \textbf{CMB + DESI DR2 + Union3} \\
\hline
$\Omega_{\mathrm{\text{cdm}}}h^2$ 
& $0.11771 \pm 0.00061$ 
& $0.11789 \pm 0.00060$
& $0.11797 \pm 0.00060$
& $0.11789 \pm 0.00061$ \\

$\Omega_{\mathrm{\text{b}}}h^2$ 
& $0.02235 \pm 0.00012$ 
& $0.02233 \pm 0.00012$
& $0.02232 \pm 0.00012$
& $0.02233 \pm 0.00012$ \\

$100\,\theta_{\mathrm{MC}}$ 
& $1.04102 \pm 0.00023$ 
& $1.04100 \pm 0.00023$
& $1.04100 \pm 0.00023$
& $1.04101 \pm 0.00023$ \\

$\tau_{\mathrm{reio}}$ 
& $0.0619_{-0.0072}^{+0.0064}$ 
& $0.0613_{-0.0073}^{+0.0063}$ 
& $0.0611 \pm 0.0067$
& $0.0615 \pm 0.0070$ \\

$n_{\mathrm{s}}$ 
& $0.9692 \pm 0.0034$ 
& $0.9687 \pm 0.0032$
& $0.9686 \pm 0.0033$
& $0.9687 \pm 0.0033$ \\

$\ln\bigl(10^{10}\,A_{\mathrm{s}}\bigr)$ 
& $3.057 ± 0.012$ 
& $3.055_{-0.013}^{+0.011}$
& $3.055 \pm 0.012$
& $3.056 \pm 0.012$ \\

\hline

$H_0 (\mathrm{km\,s^{-1}\,Mpc^{-1}})$ 
& $68.17 \pm 0.28$ 
& $68.08 \pm 0.27$
& $68.05 \pm 0.27$
& $68.08 \pm 0.28$ \\

$\Omega_{\mathrm{m}}$ 
& $0.3028 \pm 0.0036$ 
& $0.3039 \pm 0.0035$
& $0.3044 \pm 0.0035$
& $0.3039 \pm 0.0036$ \\

$\sigma_{8}$ 
& $0.8098 \pm 0.0047$ 
& $0.8152 \pm 0.0066$
& $0.8101 \pm 0.0047$
& $0.8101 \pm 0.0049$ \\

$S_{8}$ 
& $0.8136 \pm 0.0067$ 
& $0.8100_{-0.0050}^{+0.0044}$
& $0.8160 \pm 0.0066$
& $0.8153 \pm 0.0069$ \\

$r_d\text{(Mpc)}$ 
& $147.74 \pm 0.18$ 
& $147.71 \pm 0.18$
& $147.70 \pm 0.18$
& $147.71 \pm 0.18$ \\

\hline
\end{tabular}
}
\caption{This table presents the constraints on the $\Lambda$CDM model parameters, reported as mean values with their corresponding 68\% ($1\sigma$) confidence intervals, obtained from the combined analyses of DESI DR2, CMB, and SNe~Ia datasets (Pantheon$^+$, DES-Dovekie, and Union3).}\label{tab_2}
\end{table*}
\begin{table*}
\centering
\resizebox{\textwidth}{!}{%
\begin{tabular}{l@{\hspace{10pt}}c@{\hspace{10pt}}c@{\hspace{10pt}}c@{\hspace{10pt}}c}
\hline
\textbf{Parameters} 
& \textbf{CMB + DESI DR2} 
& \textbf{CMB + DESI DR2 + Pantheon$+$} 
& \textbf{CMB + DESI DR2 + DES-Dovekie} 
& \textbf{CMB + DESI DR2 + Union3} \\
\hline
$\Omega_{\mathrm{\text{cdm}}}h^2$ 
& $0.11949 \pm 0.00080$ 
& $0.11898 \pm 0.00077$
& $0.11912 \pm 0.00080$
& $0.11923 \pm 0.00079$ \\

$\Omega_{\mathrm{\text{b}}}h^2$ 
& $0.02222 \pm 0.00012$ 
& $0.02225 \pm 0.00012$
& $0.02224 \pm 0.00013$
& $0.02223 \pm 0.00012$ \\

$100\,\theta_{\mathrm{MC}}$ 
& $1.04080 \pm 0.00024$ 
& $1.04087 \pm 0.00024$
& $1.04085 \pm 0.00024$
& $1.04084 \pm 0.00024$ \\

$\tau_{\mathrm{reio}}$ 
& $0.0518 \pm 0.0068$ 
& $0.0548 \pm 0.0068$
& $0.0543 \pm 0.0069$
& $0.0533 \pm 0.0067$ \\

$n_{\mathrm{s}}$ 
& $0.9647 \pm 0.0036$ 
& $0.9659 \pm 0.0036$
& $0.9656 \pm 0.0036$
& $0.9653 \pm 0.0036$ \\

$\ln\bigl(10^{10}\,A_{\mathrm{s}}\bigr)$ 
& $3.037 \pm 0.012$ 
& $3.043 \pm 0.012$
& $3.041 \pm 0.012$
& $3.040 \pm 0.012$ \\

$w_0$ 
& $-0.43 \pm 0.21$ 
& $-0.838 \pm 0.054$
& $-0.806 \pm 0.056$
& $-0.672 \pm 0.086$ \\

$w_a$ 
& $-1.70 \pm 0.58$ 
& $-0.61_{-0.19}^{+0.21}$
& $-0.73_{-0.21}^{+0.23}$
& $-1.06_{-0.26}^{+0.30}$ \\
\hline

$H_0 (\mathrm{km\,s^{-1}\,Mpc^{-1}})$ 
& $63.7_{-2.0}^{+1.7}$ 
& $67.51 \pm 0.60$
& $67.32 \pm 0.55$
& $65.92 \pm 0.84$ \\

$\Omega_{\mathrm{m}}$ 
& $0.352 \pm 0.021$ 
& $0.3114 \pm 0.0057$
& $0.3134 \pm 0.0054$
& $0.3272 \pm 0.0086$ \\

$\sigma_{8}$ 
& $0.780 \pm 0.016$ 
& $0.8106 \pm 0.0076$
& $0.8099 \pm 0.00722$
& $0.7983 \pm 0.0088$ \\

$S_{8}$ 
& $0.844 \pm 0.011$ 
& $0.8257 \pm 0.0074$
& $0.8277 \pm 0.0075$
& $0.8335 \pm 0.0079$ \\

$r_d\text{(Mpc)}$ 
& $147.41 \pm 0.21$ 
& $147.50 \pm 0.20$
& $147.48 \pm 0.21$
& $147.46 \pm 0.20$ \\

\hline
\end{tabular}
}
\caption{This table presents the constraints on the CPL model parameters, reported as mean values with their corresponding 68\% ($1\sigma$) confidence intervals, obtained from the combined analyses of DESI DR2, CMB, and SNe~Ia datasets (Pantheon$^+$, DES-Dovekie, and Union3).}\label{tab_3}
\end{table*}
\begin{table*}
\centering
\resizebox{\textwidth}{!}{%
\begin{tabular}{l@{\hspace{10pt}}c@{\hspace{10pt}}c@{\hspace{10pt}}c@{\hspace{10pt}}c}
\hline
\textbf{Parameters} 
& \textbf{CMB + DESI DR2} 
& \textbf{CMB + DESI DR2 + Pantheon$+$} 
& \textbf{CMB + DESI DR2 + DES-Dovekie} 
& \textbf{CMB + DESI DR2 + Union3} \\
\hline
$\Omega_{\mathrm{\text{cdm}}}h^2$ 
& $0.11898 \pm 0.00076$ 
& $0.11941 \pm 0.00072$
& $0.11937 \pm 0.00071$
& $0.11897 \pm 0.00074$ \\

$\Omega_{\mathrm{\text{b}}}h^2$ 
& $0.02225 \pm 0.00012$ 
& $0.02222 \pm 0.00012$
& $0.02222 \pm 0.00012$
& $0.02225 \pm 0.00012$ \\

$100\,\theta_{\mathrm{MC}}$ 
& $1.04086 \pm 0.00024$ 
& $1.04082 \pm 0.00024$
& $1.04081 \pm 0.00024$
& $1.04087 \pm 0.00024$ \\

$\tau_{\mathrm{reio}}$ 
& $0.0548 \pm 0.0069$ 
& $0.0524 \pm 0.0066$
& $0.0528 \pm 0.0065$
& $0.0548_{-0.0071}^{+0.0063}$ \\

$n_{\mathrm{s}}$ 
& $0.9659 \pm 0.0035$ 
& $0.9648 \pm 0.0034$
& $0.9650 \pm 0.0035$
& $0.9660 \pm 0.0036$ \\

$\ln\bigl(10^{10}\,A_{\mathrm{s}}\bigr)$ 
& $3.043 \pm 0.012$ 
& $3.038 \pm 0.012$
& $3.039 \pm 0.012$
& $3.042 \pm 0.012$ \\

$B$ 
& $-0.372 \pm 0.011$ 
& $-0.3841 \pm 0.0082$
& $-0.3833 \pm 0.0077$
& $-0.3710 \pm 0.0092$ \\
\hline

$H_0 (\mathrm{km\,s^{-1}\,Mpc^{-1}})$ 
& $66.45 \pm 0.76$ 
& $67.27 \pm 0.57$
& $67.22 \pm 0.52$
& $66.37 \pm 0.63$ \\

$\Omega_{\mathrm{m}}$ 
& $0.3214 \pm 0.0070$ 
& $0.3144 \pm 0.0052$
& $0.3148 \pm 0.0048$
& $0.3221 \pm 0.0059$ \\

$\sigma_{8}$ 
& $0.8017 \pm 0.0094$ 
& $0.8108 \pm 0.0076$
& $0.8104 \pm 0.0072$
& $0.8009 \pm 0.0084$ \\

$S_{8}$ 
& $0.8298 \pm 0.0067$ 
& $0.8300 \pm 0.0067$
& $0.8301 \pm 0.0066$
& $0.8299 \pm 0.0066$ \\

$w_0$ 
& $-0.5052 \pm 0.0195$ 
& $-0.5269 \pm 0.0147$
& $-0.5254 \pm 0.0136$
& $-0.5032 \pm 0.0164$ \\

$w_{de,0}$ 
& $-0.7442 \pm 0.0221$ 
& $-0.7684 \pm 0.0166$
& $-0.7667 \pm 0.0154$
& $-0.7422 \pm 0.0186$ \\

$r_d\text{(Mpc)}$ 
& $147.50 \pm 0.20$ 
& $147.43 \pm 0.20$
& $147.44 \pm 0.19$
& $147.51 \pm 0.20$ \\
\hline

$\ln B_{\Lambda\mathrm{CDM},\,\mathrm{Anton\mbox{-}Schmidt}}$
& $-4.11$
& $-3.00$
& $-5.57$
& $-7.14$ \\

$\ln B_{\mathrm{CPL},\,\mathrm{Anton\mbox{-}Schmidt}}$
& $-7.96$
& $-9.24$
& $-10.15$
& $-9.76$ \\
\hline
\end{tabular}
}
\caption{This table presents the constraints on the Anton--Schmidt model parameters, reported as mean values with their corresponding 68\% ($1\sigma$) confidence intervals, obtained from the combined analyses of DESI DR2, CMB, and SNe~Ia datasets (Pantheon$^+$, DES-Dovekie, and Union3).}\label{tab_4}
\end{table*}
\section{Results and Conclusions}\label{sec_4}
Figs.~\ref{fig_2}, \ref{fig_3}, and \ref{fig_4} show the marginalized parameter constraints for the $\Lambda$CDM, CPL, and Anton-Schmidt models, respectively, obtained from the combined analyses of DESI DR2 + CMB, DESI DR2 + CMB + Pantheon$+$, DESI DR2 + CMB + DES-Dovekie, and DESI DR2 + CMB + Union3 datasets. The off-diagonal panels display the two-dimensional (2D) posterior contours, illustrating the correlations between different parameter combinations, where the inner and outer shaded regions correspond to the 68\% ($1\sigma$) and 95\% ($2\sigma$) confidence levels, respectively. The diagonal panels show the corresponding 1D marginalized posterior distributions of the model parameters. The corresponding mean parameter values and their associated 68\% ($1\sigma$) confidence intervals are summarized in Tables~\ref{tab_2}, \ref{tab_3}, and \ref{tab_4} for the $\Lambda$CDM, CPL, and Anton-Schmidt models, respectively, using the combined analyses of DESI DR2 + CMB, DESI DR2 + CMB + Pantheon$+$, DESI DR2 + CMB + DES-Dovekie, and DESI DR2 + CMB + Union3 datasets.

In this paper, our main goal is to compare the Anton-Schmidt model with the $\Lambda$CDM and CPL models. Therefore, we discuss only the implications of the Anton-Schmidt model, while the $\Lambda$CDM and CPL models are used as reference models. First, we discuss whether the Anton-Schmidt model is capable of alleviating the $H_0$ tension. In order to address the $H_0$ tension, the inferred value of $H_0$ in the Anton-Schmidt model should be consistent with the SH0ES measurement based on the Cepheid-calibrated Type Ia supernova distance ladder, namely $(73.04 \pm 1.04),\mathrm{km,s^{-1},Mpc^{-1}}$~\cite{riess2022comprehensive}. For the dataset combinations DESI DR2 + CMB, DESI DR2 + CMB + Pantheon$+$, DESI DR2 + CMB + DES-Dovekie, and DESI DR2 + CMB + Union3, we find deviations from the SH0ES measurement at the levels of $5.12\sigma$, $4.87\sigma$, $5.01\sigma$, and $5.49\sigma$, respectively. Therefore, the Anton-Schmidt model does not provide a significant alleviation of the $H_0$ tension within the considered observational combinations.

This can be confirmed by the inferred values of the sound horizon $r_d$. We find that the Anton-Schmidt model predicts $r_d = 147.50 \pm 0.20$, $147.43 \pm 0.20$, $147.44 \pm 0.19$, and $147.51 \pm 0.20\ \mathrm{Mpc}$ for the dataset combinations DESI DR2 + CMB, DESI DR2 + CMB + Pantheon$+$, DESI DR2 + CMB + DES-Dovekie, and DESI DR2 + CMB + Union3, respectively. These values remain very close to those predicted by the standard $\Lambda$CDM model ($r_d = 147.09 \pm 0.2$) Mpc \cite{aghanim2020planck}. It also shows that the Anton-Schmidt model modifies only the late-time evolution, while the pre-recombination expansion history remains nearly unchanged. As a consequence, in this kind of physics, one cannot reduce the sound horizon at the drag epoch, $r_d$. Therefore, even if the value of $H_0$ increases in any such model, $r_d$ remains essentially unchanged, causing the model predictions to become inconsistent with BAO measurements.

Indeed, $r_d$ and $H_0$ can be viewed as two sides of the same coin. Indeed, there have been several attempts to alleviate the $H_0$ tension. One example is the Early Dark Energy (EDE) scenario \cite{Bella:2026zuk,Poulin:2018dzj,Smith:2019ihp}, which introduces modifications to the expansion history around the redshift range $z \sim 10^3$-$10^4$. The EDE model can provide a better fit to the combined BAO + CMB observations and partially reduce the $H_0$ tension. However, it does not provide a perfect fit to the SNe~Ia data. Moreover, in such models, the reduction of $r_d$ typically requires a higher physical matter density, which in turn tends to worsen the, albeit less significant, $S_8$ tension \cite{knox2020hubble,Chaussidon:2025npr}.

The effect of the Anton--Schmidt model at late times can also be understood by analyzing the ratio $\rho_\ast/\rho_{c,0}$. Using Eq.~\eqref{eq:B_def} and the MCMC constraints reported in Table~\ref{tab_2}, we infer the characteristic density scale of the Anton--Schmidt model. We find
\[
\begin{aligned}
\frac{\rho_\ast}{\rho_{c,0}}
&=
\Omega_{\rm m,0}e^{-1/B}
\\[2mm]
&=
\begin{cases}
4.73, & \text{CMB + DESI DR2},\\
4.25, & \text{CMB + DESI DR2 + Pantheon$+$},\\
4.28, & \text{CMB + DESI DR2 + DES-Dovekie},\\
4.77, & \text{CMB + DESI DR2 + Union3}.
\end{cases}
\end{aligned}
\]
Thus, the characteristic density scale is only a few times larger than the present critical density, with $\rho_\ast/\rho_{c,0}\sim 4$-$5$. These values are smaller than the characteristic density scales commonly considered in pure logotropic-inspired models, where the reference density is usually associated with a fundamental high-energy scale. The difference originates from the different definition of the parameter $B$ adopted in the present Anton--Schmidt framework. Consequently, the quantity $\rho_\ast$ should not be interpreted as a fundamental microscopic density, but rather as an effective reference density entering the logarithmic equation of state. The inferred values, $\rho_\ast/\rho_{c,0}\sim4$--$5$, indicate that this characteristic scale is only about one order of magnitude larger than the present matter density, since $\rho_{m,0}=\Omega_{m,0}\rho_{c,0}\simeq0.31,\rho_{c,0}$, implying $\rho_\ast\simeq(13$--$15)\rho_{m,0}$. This shows that the Anton-Schmidt correction mainly influences the late-time expansion history.

A similar behaviour can also be observed in the case of the CPL model. As this model belongs to the family of dynamical dark energy models, it mainly modifies the post-recombination expansion history, while the sound horizon at recombination remains unchanged. Consequently, it cannot reduce the sound horizon after recombination and therefore remains unable to resolve the $H_0$ tension \cite{Bernal:2016gxb,Aylor:2018drw,Knox:2019rjx,Efstathiou:2021ocp,Jiang:2024xnu,Pedrotti:2025ccw}. Also, in the case of dynamical dark energy models, as preferred by the DESI DR2 \cite{karim2025desi}, we require $w_0>-1$, consequently implying $f_{\rm DE}>1$. It is known from \cite{Colgain:2025nzf} that there exists a negative correlation between $f_{\rm DE}$ and $H_0$. Indeed, in this case, the model always predicts lower values of $H_0$. In the opposite case, where $w_0<-1$ and $f_{\rm DE}<1$, the model can predict higher values of $H_0$. Yet, since $r_d$ cannot be changed, the corresponding predictions of the model become inconsistent with the BAO measurements \cite{vagnozzi2023seven,jedamzik2021reducing}.

For the Anton-Schmidt model, we obtain $\sigma_8 = 0.8017 \pm 0.0094$, $0.8108 \pm 0.0076$, $0.8104 \pm 0.0072$, and $0.8009 \pm 0.0084$, while the corresponding values of $S_8$ are $0.8298 \pm 0.0067$, $0.8300 \pm 0.0067$, $0.8301 \pm 0.0066$, and $0.8299 \pm 0.0066$ for the dataset combinations CMB + DESI DR2, CMB + DESI DR2 + Pantheon${+}$, CMB + DESI DR2 + DES-Dovekie, and CMB + DESI DR2 + Union3, respectively. For the CPL model, we obtain $\sigma_8 = 0.780 \pm 0.016$, $0.8106 \pm 0.0076$, $0.8099 \pm 0.0072$, and $0.7983 \pm 0.0088$, while the corresponding values of $S_8$ are $0.844 \pm 0.011$, $0.8257 \pm 0.0074$, $0.8277 \pm 0.0075$, and $0.8335 \pm 0.0079$ for the same dataset combinations, respectively. The corresponding values of $S_8$ remain very close to those reported by the Planck CMB observations and show a noticeable deviation from the lower values preferred by weak-lensing surveys such as KiDS, DES, and HSC~\cite{abbott2022dark,asgari2021kids}. Indeed, this suggests that the Anton-Schmidt and CPL models are not successful in resolving either the $H_0$ tension or the $S_8$ tension within the current observational scenario.

The parameter $B$ is constrained to be $-0.372 \pm 0.011$, $-0.3841 \pm 0.0082$, $-0.3833 \pm 0.0077$, and $-0.3710 \pm 0.0092$ for the dataset combinations CMB + DESI DR2, CMB + DESI DR2 + Pantheon$+$, CMB + DESI DR2 + DES-Dovekie, and CMB + DESI DR2 + Union3, respectively. These negative values are consistent with the theoretical expectation of the Anton-Schmidt framework, where $B<0$ naturally arises when $\rho_\ast>\rho_{m,0}$ and gives rise to an effective negative pressure responsible for the late-time accelerated expansion of the Universe. Moreover, since $B$ determines the characteristic density scale through Eq.~\eqref{eq:B_def}, the inferred values imply that the Anton-Schmidt correction becomes important only at relatively low cosmic densities, supporting its interpretation as a late-time dynamical dark-energy model. The corresponding present-day dark-energy equation-of-state parameter is constrained to be $w_{de,0} = -0.7442 \pm 0.0221$, $-0.7684 \pm 0.0166$, $-0.7667 \pm 0.0154$, and $-0.7422 \pm 0.0186$ for the same dataset combinations, respectively. These values lie within the interval $-1 < w_{de,0} < -1/3$, indicating that the Anton-Schmidt model shows a quintessence-like dark-energy behavior.

Similarly, for the CPL model, we obtain $w_0 = -0.43 \pm 0.21$, $-0.838 \pm 0.054$, $-0.806 \pm 0.056$, and $-0.672 \pm 0.086$ for the dataset combinations CMB + DESI DR2, CMB + DESI DR2 + Pantheon$+$, CMB + DESI DR2 + DES-Dovekie, and CMB + DESI DR2 + Union3, respectively. These values lie within the interval $-1 < w_0 < -1/3$, indicating that the CPL model also shows a quintessence-like dark-energy behavior across all dataset combinations.

The total equation-of-state parameter at the present epoch is constrained to be $w_{de},0 = -0.5052 \pm 0.0195$, $-0.5269 \pm 0.0147$, $-0.5254 \pm 0.0136$, and $-0.5032 \pm 0.0164$ for the corresponding dataset combinations. Since all values satisfy $w_0 < -1/3$, the total cosmic fluid in the Anton-Schmidt model remains capable of driving the current accelerated expansion of the Universe, while showing only a mild deviation from the standard $\Lambda$CDM scenario.

We first discuss the possible mathematical singularity appearing in the dark-energy equation-of-state parameter of the Anton-Schmidt model. From Eq.~\eqref{eq:AS_wde}, the dark-energy equation-of-state formally diverges when $1-3B\ln a=0$. This divergence arises from the denominator of the dark-energy equation of state and corresponds to the epoch at which the dark-energy density vanishes while the pressure remains finite, leading to a breakdown of the effective fluid decomposition rather than a true spacetime singularity.

In Fig.~\ref{fig_5}, we show the evolution of the dark-energy equation-of-state parameter as a function of redshift for the Anton-Schmidt model (left panel) and the CPL model (right panel). As discussed above, the exact expression given by Eq.~\eqref{eq:AS_wde} diverges when $1-3B\ln a=0$. Therefore, to describe the cosmological evolution of the Anton--Schmidt model, we use the approximate expression given by Eq.~\eqref{eq:AS_wde_approx}. This approximation remains well behaved over the relevant redshift range and captures the dynamical dark-energy behavior of the Anton-Schmidt model.

In the case of the Anton-Schmidt model, we find a phantom crossing at $z \approx 0.5$, where the dark-energy equation-of-state parameter evolves from the phantom regime, $w_{\rm de} < -1$, at high redshifts to the quintessence regime, $w_{\rm de} > -1$, at low redshifts, including the present epoch, during the evolution of the Universe. A similar behavior can be observed for the CPL model (right panel), where the dark-energy equation-of-state also shows a phantom-crossing evolution from the phantom regime to the quintessence regime. Indeed, in both cases, the crossing occurs at approximately the same redshift, $z \approx 0.5$. This kind of evolution, in which the dark-energy equation-of-state parameter evolves from the phantom regime in the past to the quintessence regime at late times (including the present epoch), is known as the \emph{Quintom-B} scenario \cite{Feng:2004ad}. We emphasize that the approximate expression given by Eq.~\eqref{eq:AS_wde_approx} is used only to shows the background evolution of the dark-energy equation-of-state parameter and the phantom-crossing behavior. The CMB anisotropy and matter power spectra are computed using the modified CAMB Boltzmann solver, where dark-energy perturbations are evolved within the Parametrized Post-Friedmann (PPF) framework.

Now, we use the Bayesian evidence as a statistical test to compare the Anton-Schmidt model with the $\Lambda$CDM and CPL models. We find $\ln \mathcal{B}_{\Lambda{\rm CDM},{\rm Anton\mbox{-}Schmidt}}=-4.11$, $-3.00$, $-5.57$, and $-7.14$ for the dataset combinations CMB + DESI DR2, CMB + DESI DR2 + Pantheon$^+$, CMB + DESI DR2 + DES-Dovekie, and CMB + DESI DR2 + Union3, respectively. Similarly, we obtain $\ln \mathcal{B}_{{\rm CPL},{\rm Anton\mbox{-}Schmidt}}=-7.96$, $-9.24$, $-10.15$, and $-9.76$ for the same dataset combinations, respectively. These negative values indicate that the Bayesian evidence favors the Anton-Schmidt model over both the $\Lambda$CDM and CPL models. According to the revised Jeffreys' scale, the preference over $\Lambda$CDM is moderate for the CMB + DESI DR2 and CMB + DESI DR2 + Pantheon$^+$ combinations, and strong for the CMB + DESI DR2 + DES-Dovekie and CMB + DESI DR2 + Union3 combinations. In comparison, the preference over the CPL model is strong for all dataset combinations, except for CMB + DESI DR2 + DES-Dovekie, where it reaches the decisive level.

In summary, we performed a comprehensive observational analysis of the Anton-Schmidt dark energy model using the latest DESI DR2 BAO measurements in combination with Planck CMB and three independent Type Ia supernova compilations (Pantheon$+$, DES-Dovekie, and Union3). We compared the Anton-Schmidt model with the standard $\Lambda$CDM and CPL parametrizations through MCMC parameter estimation, cosmological observables, and Bayesian model selection. Our results show that the Anton-Schmidt model does not provide a significant alleviation of either the $H_0$ or the $S_8$ tensions, while the sound horizon at the drag epoch remains essentially unchanged. This is because the characteristic density scale satisfies $\rho_\ast/\rho_{c,0}\sim4$--$5$, indicating that the Anton-Schmidt correction becomes relevant only at late times and therefore mainly affects the recent expansion history of the Universe. Consequently, the model modifies only the post-recombination expansion history while leaving the sound horizon at recombination and the early-Universe physics essentially unchanged, a behavior that is expected for late-time dynamical dark-energy models. Nevertheless, the model predicts a quintessence-like present-day dark-energy equation of state and shows a Quintom-B evolution, with the dark-energy equation-of-state parameter crossing the phantom divide at approximately $z\approx0.5$.

From the Bayesian model comparison, however, the Anton-Schmidt model is statistically preferred over both the $\Lambda$CDM and CPL models. According to the revised Jeffreys' scale, the preference over $\Lambda$CDM ranges from moderate to strong depending on the dataset combination, whereas the preference over the CPL parametrization is consistently strong and becomes decisive for the CMB + DESI DR2 + DES-Dovekie dataset. These results suggest that the logarithmic equation of state of the Anton-Schmidt model provides a competitive phenomenological description of current cosmological observations. Nevertheless, the present analysis is restricted to the constant-$n$ ($n=-1$) realization of the Anton-Schmidt model, which should be regarded as an effective late-time description. Since the Anton-Schmidt parameter is expected to acquire a temperature dependence through the Grüneisen parameter, a complete cosmological analysis would require implementing the full evolution of $n(T)$ in both the background and perturbation equations of the modified CAMB Boltzmann solver, followed by a new cosmological parameter estimation.

A natural next step will be to confront the model with forthcoming Stage-IV large-scale structure observations, including the LSST Year 1 and Roman Space Telescope $3\times2$-point analyses, as well as DES-Y3$\times$Planck measurements, using the  Cobaya-Cosmolike Joint Architecture (CoCoA) likelihood framework \cite{Krause:2016jvl}. Such analyses will provide substantially more stringent constraints on the Anton-Schmidt model and offer a robust test of its viability on both background and perturbation scales. Extending the analysis to the full temperature-dependent Anton-Schmidt scenario together with these next-generation cosmological datasets is left for future work.

\section*{Acknowledgements}
SC acknowledges the Istituto Nazionale di Fisica Nucleare (INFN) Sez. di Napoli,  Iniziative Specifiche QGSKY and MoonLight-2  and the Istituto Nazionale di Alta Matematica (INdAM), gruppo GNFM, for the support. This paper is based upon work from COST Action CA21136 -- Addressing observational tensions in cosmology with systematics and fundamental physics (CosmoVerse), supported by COST (European Cooperation in Science and Technology).

\bibliographystyle{elsarticle-num}
\bibliography{mybib.bib}

\end{document}